\documentclass[conference,final]{IEEEtran}

\IEEEoverridecommandlockouts

\usepackage[utf8]{inputenc}
\usepackage{bm}
\usepackage{amsfonts,amsmath,amssymb}
\usepackage{graphicx} 
\usepackage{soul}
\usepackage{siunitx}
\usepackage{cite}
\usepackage{balance}
\usepackage[caption=false]{subfig}
\usepackage{enumitem}
\usepackage{comment}

\usepackage{multirow}
\usepackage[table,xcdraw]{xcolor}

\usepackage{pgfplots}
\pgfplotsset{compat=1.18}
\usepackage{tikz}
\usepackage{pgfplotstable}
\usetikzlibrary{plotmarks, decorations, backgrounds, spy, arrows, fadings, patterns, shadows.blur, shapes}
\definecolor{col3}{RGB}{8,104,172}
\definecolor{col2}{RGB}{145,207,96}
\definecolor{col1}{RGB}{215,48,39}
\definecolor{purp}{RGB}{136,86,167}

\usepackage{dsfont}

\tikzset{new spy style/.style={spy scope={magnification=4,size=2cm, connect spies,every spy on node/.style={rectangle,draw,}}},	every spy in node/.style={draw,rectangle,}}

\usepackage[acronym,shortcuts]{glossaries}
\newacronym{awgn}{AWGN}{additive white Gaussian noise}

\newacronym{bi}{BI}{Bellman iterative solution}

\newacronym{cr}{CR}{challenge-response}
\newacronym{csi}{CSI}{shannel state information}

\newacronym{det}{DET}{detection error tradeoff}

\newacronym{fa}{FA}{false alarm}
\newacronym{glrt}{GLRT}{generalized likelihood ratio test}

\newacronym{iod}{IoD}{Internet-of-Drones}

\newacronym{los}{LOS}{line-of-sight}
\newacronym{lrt}{LRT}{likelihood ratio test}
\newacronym{lt}{LT}{likelihood test}

\newacronym{md}{MD}{missed detection}
\newacronym{mdp}{MDP}{Markov decision process}

\newacronym{ntn}{NTN}{non-terrestrial networks}

\newacronym{pg}{PG}{purely greedy solution}
\newacronym{pla}{PLA}{physical layer authentication}

\newacronym{std}{STD}{standard deviation}
\newacronym{stdsol}{STD}{standard deviation-based solution}

\DeclareMathOperator*{\argmax}{arg\,max} 
\DeclareMathOperator*{\argmin}{arg\,min} 
 \newlength\fwidth
\newlength\fheight
\DeclareRobustCommand{\IEEEauthorrefmark}[1]{\smash{\textsuperscript{\footnotesize #1}}}

\title{Energy-Based Optimization of Physical-Layer Challenge-Response Authentication with Drones\thanks{This work was partially funded by the European Commission through the Horizon Europe/JU SNS project ROBUST-6G (Grant Agreement no. 101139068). Corresponding author: Francesco Ardizzon, email: francesco.ardizzon@unipd.it.}}
\author{%
    \IEEEauthorblockN{Francesco Ardizzon\IEEEauthorrefmark{1}, Damiano Salvaterra\IEEEauthorrefmark{1},  Mattia Piana\IEEEauthorrefmark{1}, and Stefano Tomasin\IEEEauthorrefmark{1}\IEEEauthorrefmark{2}}\vspace{1mm}%
    \small%
    \IEEEauthorblockA{\IEEEauthorrefmark{1}Department of Information Engineering, University of Padova, Italy}%
    \IEEEauthorblockA{\IEEEauthorrefmark{2}Department of Mathematics, University of Padova, and CNIT, Parma, Italy}%
}

\begin{document}

\maketitle

\sloppy 

\begin{abstract}
Drones are expected to be used for many tasks in the future and require secure communication protocols. In this work, we propose a novel \ac{pla}-based \ac{cr} protocol in which a drone Bob authenticates the sender (either on the ground or air) by exploiting his prior knowledge of the wireless channel statistic (fading, path loss, and shadowing). In particular, Bob will move to a set of positions in the space, and by estimating the attenuations of the received signals he will authenticate the sender. We take into account the energy consumption in the design and provide three solutions: a \ac{pg}, an optimal \ac{bi}, and a heuristic solution based on the evaluation of the standard deviation of the attenuations in the space. Finally, we demonstrate the effectiveness of our approach through numerical simulations.
\end{abstract}

\glsresetall
 
\section{Introduction}\label{sec:intro}

Nowadays, drones are being used for various tasks such as precision agriculture and disaster management. Moreover, the integration of drones in machine-to-machine communication is helpful in a variety of contexts, such as the \ac{iod}\cite{Alam22} and \ac{ntn} \cite{Lin21}. However, these capabilities also make drones possible targets of attacks, including, for instance, spoofing to disrupt the drone's navigation system \cite{ceccato21,michieletto22}, and jamming as a denial of service. Security threats and countermeasures for drone communications have recently been reviewed in \cite{Adil23systematic}.  %in \cite{Hassija21Fast} and \cite{Adil23systematic}.

This paper addresses the problem of drone authentication, where a drone, Bob, communicates with a transmitter Alice on the ground, while a third {\em intruder} agent, Trudy, aims to impersonate Alice and send fake malicious messages to Bob. The goal of Bob is to verify the authenticity of the sender and distinguish Alice from Trudy.

A classical authentication strategy is the (cryptography-based) \ac{cr} protocol, which is based on a secret key shared between Alice and Bob \cite[Sec. 13.5]{gupta2014cryptography}. In this protocol, the verifier sends a request, called a \emph{challenge}, which only a legitimate user with a valid key can correctly answer. In this paper, we consider \ac{pla}-based \ac{cr} solutions first introduced in \cite{CR-PLA:PCC}. \ac{cr}-\ac{pla} is based on \emph{partially controllable channels} as the verifier Bob authenticates Alice by manipulating the channel and verifying that the received signal is consistent with the expected change. The change induced by Bob corresponds to the {\em challenge} of crypto-based \ac{cr}. On the other hand, since the channel change is randomly decided by the verifier, it is difficult for Trudy to predict Bob's manipulations and the resulting channel.  The advantages of such a class of protocols are detailed in \cite{CR-PLA:PCC}. 
 
In this paper, we propose a \ac{cr}-\ac{pla} protocol for drone authentication, where the controllable parameter used as the challenge is the drone's position which impacts the path loss of the resulting wireless link. Even a sophisticated Trudy (which can freely tune its transmission power and pre-compensate the Trudy-Bob channel) that does not know the current drone's position has high uncertainty on the legitimate channel, strengthening the authentication procedure. In this scenario, the attacker's probability of success depends on the variability of the attenuation over the space of movement of Bob. Thus, we use the shadowing effect as a source of randomness. In particular, Bob checks whether the received response matches the distribution of attenuation due to shadowing given his position.
Since more positions can be associated with the same (quantized) path loss and shadowing (i.e., the challenge), we propose an energy-saving \ac{cr}-\ac{pla} strategy that minimizes the long-term energy expended by Bob while ensuring a target security level.

In \cite{CR-PLA:DN}, the authors proposed a \ac{cr}-\ac{pla} protocol for drone communications. However, the channel was only partially modeled (e.g., no shadowing was considered), and the problem of energy minimization was not addressed. 

In detail, the contributions of the paper are the following:
\begin{itemize}
    \item a novel \ac{cr}-\ac{pla} protocol for drone authentication using the channel shadowing.
    \item a verification procedure that is resilient to noise, e.g., due to fading and signal processing errors.
    \item a policy to minimize the long-term energy spent by the drone without compromising security. In detail, we propose a \ac{pg}, an optimal \ac{bi}, and a heuristic solution based on the \ac{std} of attenuations in the space. 
\end{itemize}
 
The rest of the paper is organized as follows. Section~\ref{sec:sysModel} introduces the system model. Section~\ref{sec:proto} analyzes the security of the proposed \ac{cr}-\ac{pla} protocol, while the policy designs for energy minimization are discussed in Section~\ref{secopt}. In Section~\ref{sec:results} we show the numerical results. Finally, Section~\ref{sec:conclusion} draws the conclusions.

\section{System Model}\label{sec:sysModel}

\begin{figure}[b]
    \centering
    %\resizebox{\columnwidth}{!}{\input{figures/scheme}}            
    \input{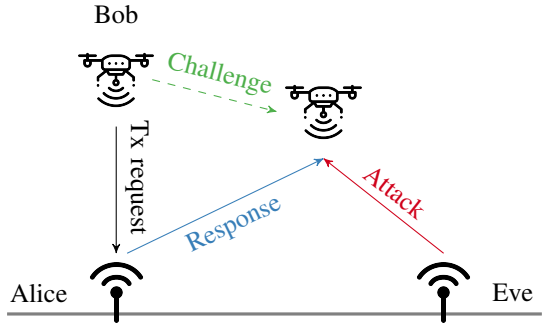}
    \caption{Example of \ac{cr} protocol.}
    \label{fig:sketch}
\end{figure}

We consider the scenario of Fig.~\ref{fig:sketch}, where a drone (agent Bob), is communicating with a static (ground or aerial) device (agent Alice) while moving on a gridded region of points $\mathcal{X}= \{ \bar{\bm{x}}_1, \ldots, \bar{\bm{x}}_N\}$, where $\bar{\bm{x}}_n$ is the coordinate vector of position $n$. The drone receives messages from the ground device, including those for navigation and the instructions required by the mission (e.g., about when taking pictures). We design an authentication mechanism to ensure that the drone processes only messages coming from Alice, rather than from one impersonating her, Trudy. In turn, Trudy aims to impersonate Alice by sending messages to Bob purposely designed to be confused with those of Alice. Such messages aim for example to detour the drone from its designed route.

In the next part of this Section, we describe the channel model, the channel estimation, the threat model, and the model of drone power consumption. 

\subsection{Channel Model and Estimation}
Transmissions are narrowband and channel gains are the result of (free space) path-loss, shadowing, and fading phenomena, which properly capture the main components of wireless propagation. As we are considering shadowing and path-loss as authentication features, our solution is independent of the number of antennas of each device.

As described in the next Section, Bob needs to estimate the channel over which the receiver message went through for \ac{cr}-\ac{pla}. Several well-known strategies can be used for such a task, e.g., the least squares estimation. In particular, from the received signal Bob extracts the attenuation measured in \si{\decibel} \cite{ACSA}
\begin{equation}\label{eq:shadowed_pl}
    a = \underbrace{a_\mathrm{PL}+a_\mathrm{SH}}_{\eta} +  {a}_\mathrm{FD},
\end{equation}
which includes the free space path-loss attenuation $a_\mathrm{PL}$ (modeled by the Friis formula), the shadowing attenuation $a_\mathrm{SH}$, and fading attenuation $a_\mathrm{FD}$.
%In \eqref{eq:shadowed_pl}, the first term $\hat{a}$ includes the contributions of pathloss and shadowing, while ${a}_\mathrm{FD}$ is due to fading only. 
We also assume the transmit power of Alice and antenna gains of the legitimate agents are publicly known, and thus they can be ignored in our model.% and not indicated in \eqref{eq:shadowed_pl}. 

%The value measured in \eqref{eq:shadowed_pl} will be the input of the \ac{cr}-{pla} to be tested.
\begin{comment}
    
\paragraph*{Path-loss Term} Let $P_\mathrm{Tx}$ and $P_\mathrm{Rx}$ are the transmit and receive power, respectively, $\alpha$ is the path-loss coefficient, $d$ is the transmitter-receiver distance (in km), $f_0$ is the carrier frequency (in \si{\mega\hertz}), and $G_\mathrm{Tx}$ and $G_\mathrm{Rx}$ are the transmit and receive antenna gains (in dB), respectively. Considering a deterministic ray model with \ac{los} transmissions, the total attenuation due to path loss at the receiver is modeled by the Friis formula as 
\begin{equation}\label{eq:attenuation}
\begin{split}
    a_\mathrm{PL} &= 10\log_{10}{\frac{P_\mathrm{Tx}}{P_\mathrm{Rx}}}   = \\
     &=32.4+10\alpha\log_{10}{d}+20\log_{10}{f_0} - G_\mathrm{Tx}- G_\mathrm{Rx}.
\end{split}
\end{equation}
In the following, we will assume that the antenna gains of Alice and Bob are known to all the agents (including Trudy), as they can be obtained from datasheets of the devices.
\end{comment}

Shadowing includes the effects of obstacles placed between the transmitter and receiver, thus $a_{\rm SH} \sim \mathcal{N}(0,\,\sigma^{2}_\mathrm{SH})$ has a normal distribution (in \si{\decibel} scale) with standard deviation $\sigma_\mathrm{SH}$. Typical values are $\sigma_\mathrm{SH} \in [4, 12]$ dB.
The shadowing term depends on the location of the transmitter and the receiver, and channel gains of couples of transceivers in proximity have a high correlation. To model such a phenomenon, we resort to the well-known Gudmundson \cite{ACSA} model of the correlation of the shadowing components for two receivers at a distance $\Delta$ 
\begin{equation}\label{eq:gudmundson}
r_\mathrm{SH}(\Delta)=\sigma^{2}_\mathrm{SH}\exp\left({-\frac{\Delta}{D_{\rm coh}}}\right), 
\end{equation}
%where $D_{\rm coh}$ is the coherence distance computed as \begin{equation}\label{eq:coh_d}
%    D_{\rm coh} = k\lambda = k\frac{c}{f_0}, 
%\end{equation}
where $D_{\rm coh}$ is the coherence distance. Typical values of the coherence distance range from $3\lambda$ to $20\lambda$ (where $\lambda$ is the wavelength) in an urban area with many obstacles and a flat rural area~\cite{ACSA}, respectively.

Fading accounts for shorter channel variations and depends on Doppler spread and multipath. As it is hardly predictable, it cannot be exploited in our authentication scheme. Shadowing instead is shown to have a higher coherence distance. Moreover, it varies only slowly over time (i.e., slow fading). Thus, we will exploit the shadowing for the \ac{cr} protocol while considering fading as estimation noise. 

%In particular, we assume that for a short time period, Bob has traveled around an area $\mathcal{A}$ \hl{?} and collected a set of received attenuations and was able to reconstruct a partial attenuation map, which contains the set of shadowing and free space path loss measured for each position $\bm{x}\in\mathcal{A}$. \hl{FA: mettere pi\'u details?}

\subsection{Threat Model}
Concerning the attacker Trudy, we assume that she can control her transmitting signal to induce any desired attenuation at Bob. On the other hand, we assume Bob's position to be unknown to her until signal reception, i.e., the spoofing signal does not depend on Bob's position. 

\subsection{Power Consumption for Movements}
Since the protocol requires the drone to move, we define $\epsilon(\bm{x},\bm{x}') $ as the energy needed to move from position $\bm{x}$ to position $\bm{x}'$.
Following the results in \cite{Abeywickrama18}, the energy consumption (in Joul) for both vertical and horizontal movement follows the model
\begin{equation}\label{eq:energy}
    \epsilon(\bm{x},\bm{x}') = \alpha_1\, \frac{\|\bm{x} - \bm{x}' \|}{V} - \alpha_0\;,
\end{equation}
where $V$ is the (constant) drone velocity. For instance, taking $\alpha_1 \approx \SI{308.71}{\joule/\second}$ and $\alpha_0 \approx \SI{0.85}{\joule}$, \eqref{eq:energy} fits the energy consumption when performing an horizontal movement.

\section{Challenge-Response Authentication \\ at The Physical Layer}\label{sec:proto}
Bob aims to distinguish if the received messages come from Alice or Trudy. Additionally, as we work with an energy-limited device, we aim to minimize Bob's movements to lower the energy consumption in the long run.

We propose an authentication mechanism based on the attenuation measured by Bob on the received message compared to the one expected according to the current positions of Alice and Bob. In particular, we let Bob choose a position $\bm{x}$, and such choice is considered to be the challenge in a \ac{cr} protocol. In turn, the attenuation measured when in such a position is the response. The position and thus the challenge shall be chosen taking into account security, therefore it has to guarantee low \ac{fa} and \ac{md} probabilities when testing the authenticity of the response while keeping the energy consumption as low as possible. The position is chosen also to minimize the movements, thus keeping under control the drone power consumption.

We assume that Bob has a {\em map} of quantized attenuations over the grid $\mathcal X$. In particular, indicating with ${\tt q}(\cdot)$ the quantization function, we have 
$\hat{a}(\bar{\bm{x}}_n) = {\tt q}(\eta(\bar{\bm{x}}_n))$, where $\eta$ is defined in \eqref{eq:shadowed_pl} and includes only free-space path loss and shadowing, obtained by averaging out fading over multiple measurements. The set of the  {\em unique} quantized attenuations is therefore $\mathcal{A} = \{a: \exists\,  n: \hat{a}(\bar{\bm{x}}_n)=a\}$. Moreover, for any attenuation $a' \in \mathcal{A}$, let us define the set of positions all having the same attenuation as
\begin{equation}
    \mathcal{X}_{a'} \triangleq \left\{\bm{x} \in{\mathcal{X}} \big|\hat{a}(\bm{x}) = a' \right\}\;.
\end{equation}

The proposed authentication mechanism distinguishes between Alice's and Trudy's signals based on the prior knowledge of such a map.
% Additionally, the map has an intrinsic redundancy, as the same attenuation, and thus the same response may be measured from multiple positions. Thus, the challenger has an additional degree of freedom. We exploit such a degree of freedom to minimize the energy consumption of the drone, i.e., finding the set of positions that allows the drone to pose challenges with the highest security level, while limiting the energy consumption.
%We assume the verifier to be provided with a dataset containing responses from different positions over the area in which the drone may move. This dataset will be used to both choose the challenge and verify the response, thus it needs to be authentic.

% Given an attenuation map $\{a(\bm{x})\}$ from \eqref{eq:att_MAP} and the current Bob position $\bm{x}_\mathrm{B}$, we aim to design a \ac{cr}-\ac{pla} protocol for drone authentication, allowing Bob to distinguish Alice from the attacker Trudy, who is trying instead to impersonate Alice.
% Due to the sampling of the 2D area, the set of possible attenuations is finite. Thus, Bob can collect all the possible attenuation values in the set 
% \begin{equation}
%     {\mathcal{A}} \triangleq \left\{ \tilde{a} \big| \exists \, x \in \mathcal{X}\, \mbox{s.t.}  \, a(\bm{x}) = \tilde{a} \right\}\;.
% \end{equation}

In the following, we assume that a message is transmitted every integer time $t=0, 1, ...$ only to simplify notation. In practice, messages do not need to be sent at regular time.

The proposed \ac{cr}-\ac{pla}-based algorithm is composed of the following steps, also shown in Fig. \ref{fig:sketch}. At time $t_0$ the following occur
\begin{enumerate}[start=0]
    \item \emph{Channel Measurements}: Bob moves on positions in $\mathcal{X}$ while Alice transmits several pilot signals. These transmissions are authenticated by higher-layer security mechanisms, Bob uses them to obtain $\mathcal{A}$ and $\mathcal X_a$, $a \in \mathcal A$.
\end{enumerate}
Next, when Alice wants to transmit a message at time $t>t_0$,
\begin{enumerate}
    \item \emph{Challenge}: Alice transmits in broadcast a transmission request to Bob (Tx request in Fig. \ref{fig:sketch}), who chooses uniformly at random an attenuation in $\mathcal A$, i.e., $a_t^\star\sim\mathcal{U}(\mathcal{A})$. Then Bob moves to a position $\bm{x}^\star$, for which $\hat{a}(\bm{x}^\star) = a_t^\star$. The selection of the position is discussed in Section~\ref{secopt}.
    \item \emph{Response}: Alice transmits her message, including pilot signals to Bob.
    \item \emph{Verification}: Upon reception, Bob estimates the  attenuation of the channel over which the message went, $\tilde{a}_t$ and if it statistically matches $a_t^\star$, the message is considered authentic otherwise is labeled as fake. The test procedure is presented in Section~\ref{sec:verificationProc}. 
\end{enumerate}

We remark that this procedure does not make use of pre-shared keys, except during step 0. This is common also to the tag-based \ac{pla} mechanisms where it is assumed either that it is possible to perform a secure data collection, or as mentioned before, that the verifier knows the measurement distribution in advance. 

Bob chooses $a_t^\star$ to be picked from $\mathcal{A}$ using a uniform distribution: while several different choices are possible, this specific distribution will force Trudy to use a random guessing strategy, as we will detail in the next Section.

\subsection{Authentication Verification Procedure}\label{sec:verificationProc}
In Step 1, after the transmission request, Bob will move towards ${\bm{x}}^\star$, stop, and wait for the signal from Alice. 

In Step 3, due to the fading, hardware imperfections, and errors introduced by the signal processing, Bob will measure an attenuation $\tilde{a}_t$ instead of $a^\star_t$.
We frame the problem of checking whether $\tilde{a}_t$ is a noise-corrupted version of $a^\star_t$ as a binary hypothesis testing. We consider $\mathcal{H}_0$ as the legitimate (null) hypothesis (see \eqref{eq:shadowed_pl}), and $\mathcal{H}_1$ as the under-attack hypothesis, i.e.  
\begin{equation}\label{eq:hptest}
    \tilde{a}_t = \begin{cases}
        a^\star_t + a_{\rm FD} \quad &\mbox{if } \mathcal{H}_0,\\
        b_t\quad &\mbox{if } \mathcal{H}_1\,,
    \end{cases}
\end{equation}
where $a_{\rm FD} \sim \mathcal{E}(1)$ is the exponential random variable (with parameter 1) describing the fading and $b_t$ is the attenuation measured when under attack.~\footnote{Note that in \eqref{eq:hptest} we ignored the effects of quantization, considering it negligible when compared to fading.}  Notice that we are considering a worst-case scenario where Trudy can pre-compensate the channel to account for such noise and that when under attack, only Trudy's signal is received and Alice causes no interference.

Bob then checks whether $\tilde{a}_t$ matches the statistic of the chosen challenge $a^\star_t$. 
In binary hypothesis testing, the optimal strategy is to use the \ac{lrt}. Still, this test would require the knowledge of the attenuation when under attack, which is unknown a priori. 
Thus we resort instead to the \ac{lt}, and defining $p({\tilde{a}}_t|a^\star_t, \mathcal{H}_0)$ the PDF of $\tilde{a}_t$ under the hypothesis $\mathcal{H}_0$ the \ac{lt} is 
\begin{equation}\label{eq:GLRT}
\begin{split}
      \mbox{LT}(\tilde{a}_t,\varphi ) : & p({\tilde{a}_t}|a^\star_t, \mathcal{H}_0) \lessgtr \varphi   \Longleftrightarrow \\ 
        &\mathds{1}(\tilde{a}_t - a^*_t)  \exp[-(\tilde{a}_t - a^*_t)]  
      \lessgtr \varphi \,,
\end{split}
\end{equation}
where $\mathds{1}(\cdot)$ is the Heaviside step function and $\varphi$ is a threshold chosen by Bob to meet a predefined \ac{fa} probability $P_{\rm fa}$ Note that only inputs obtaining outputs greater than $\varphi$ in \eqref{eq:GLRT} are considered authentic.  
Even if it has no general optimality guarantee, this test is often used in practice as it does not require any knowledge about the strategy employed by the attacker.

The \ac{lt} \eqref{eq:GLRT} can be interpreted as a test of distance in the probability space. In particular, $\varphi$ identifies an interval $\mathcal{I}$ around  $a^\star_t$: only values falling inside $\mathcal{I}$, i.e., close to $a^\star_t$ are considered to be authentic.
Then, a \ac{fa} occurs when a legitimate $\tilde{a}_t$ falls outside of $\mathcal{I}$, and, conversely, for a fixed \ac{fa} probability, the interval size must be (by inverting the cumulative distribution function of the exponential random variable)
% \begin{equation}
% |\mathcal{I}| = 2\sigma_w \ {Q}^{-1} \left( \frac{p_\mathrm{fa}}{2} \right)\:,
% \end{equation}
\begin{equation}
|\mathcal{I}| = -\ln P_{\rm fa}\:.
\end{equation}
%where $Q$ is the tail distribution function of the (standard) Gaussian distribution.
On the other hand, a \ac{md} occurs when the attacker manages to pick a value $b_t$ falling inside $\mathcal{I}$, thus close enough to $a^\star_t$. Even a sophisticated attacker able to compensate the Trudy-Alice channel can only perform a random guessing attack $b_t \sim \mathcal{U}(\mathcal{A})$.
Hence, the \ac{md} probability is due to the difference between the uniform random variables $b_t$ and $a^\star_t$ defined over the same domain, which results in a triangular random varible, therefore the \ac{md} probability is
% \begin{equation}\label{eq:pmd}
%     p_\mathrm{md} = \frac{|\mathcal{I}|}{r} = \frac{2\sigma_w}{r} \ {Q}^{-1} \left( \frac{p_\mathrm{fa}}{2} \right)\,,
% \end{equation}
\begin{equation}\label{eq:pmd}
    P_\mathrm{md} = \frac{1}{2} - \frac{(r + \ln P_{\rm fa})^2}{2r^2}\,,
\end{equation}
where $r = \max_n \hat{a}(\bar{\bm{x}}_n)  - \min_n  \hat{a}(\bar{\bm{x}}_n)$.
%We remark that the choice of the path is arbitrary and it could be possible to choose a different path longer than the shortest one, with no impact on the security performance of the scheme. The advantage of choosing the shortest path is that it allows us to find a tradeoff between security and energy consumption, as it reduces the average distance between the drone and the target ${\bm{x}}^\star$. \hl{FA: questo \'e intuitivo a livello di media, ma non avrebbe pi\'u senso scegliere \mbox{$\bm{x}^\star$} come la posizione  pi\'u vicina a quella del drone?}

%Notice, the highest is the sampling granularity of $\mathcal{X}$, the  more diverse map $\{a(\bm{x})\}$ becomes. That in turn, increases the number of possible challenges that be posed by Bob, enhancing the security of the scheme. Alternatively, it could be also possible to interpolate $a(\bm{x})$ thus picking a position instead not explored in Step 1. 

\section{Energy-Aware Mechanism Optimization}\label{secopt}

In this section, we discuss the choice of the position $\bm{x}^\star$ selected by Bob for \ac{cr}-\ac{pla}, taking into account both security (in terms of \ac{fa} and \ac{md} probabilities) and energy consumption for the drone movement. In particular, assuming that we run the \ac{cr}-\ac{pla} many times and that at $t=0$ the drone starts from a random initial position in $\mathcal{X}$, we design a \emph{policy} to choose the next position, which minimizes the overall energy spent at each time $t> 0$.

As previously pointed out, the protocol security is based on the statistic of the selected attenuation $a^\star_t$ and not on the actual Bob position. However, many positions may correspond to the same attenuation if the grid $\mathcal{X}$ is dense enough. Hence,  this opportunity can be exploited to minimize the energy consumption of the drone using the \ac{cr}-\ac{pla} protocol.

%For simplicity, instead of considering the whole $ \tilde{\mathcal{A}} $ space, we uniformly sample it, considering only a finite number of attenuations, $|\mathcal{A}| =|\mathcal{A}|$, collected in $ \mathcal{A}$. 
%we map to each $a^\star_{i}$ value a set of candidate target position $\mathcal{X}_{a_n}$ such that 

%Given the map containing the previously measured, first, Alice computes the values $a_\mathrm{min}$ and  $a_\mathrm{max}$, as 
%\begin{subequations}
%\begin{align}
%    a_{\rm max} = \underset{\bm{x} \in \mathcal{S}}{\max}\:\: a{(\bm{x})} \\
%    a_{\rm min}= \underset{\bm{x} \in \mathcal{S}}{\min}\:\: a{(\bm{x})}\:.
%\end{align}
%\end{subequations}
%Next, it considers the shortest path including the drone's current position, $a_\mathrm{min}$, and  $a_\mathrm{max}$. Hence, any position $\bm{x}$ included in this path will have attenuation $a(\bm{x}) \in [ a_\mathrm{min}, a_\mathrm{max}] $.
%Assuming the grid is dense enough, and thanks to the quantization of the attenuation values, there may be more than one position $\bm{x}^\star$ such that $a(\bm{x}^\star) = a^\star$. 

% Moreover, the collection of such sets forms a partition of $\mathcal{X}$, i.e., 
% \begin{equation}
%     \bigcup_{a\in \mathcal{A}} \mathcal{X}_{a} = \mathcal{X}\; ,\quad \mbox{and}\quad \mathcal{X}_{a_1}\cap \mathcal{X}_{a_2} = \emptyset\,, \; \forall a_1\neq a_2\,.
% \end{equation}
%
We formalize the problem by modeling it as a \ac{mdp}. 
In detail, we introduce the {\em state at step $t$} as $s_t \triangleq \{\bm{x}_t,a^\star_t\}$, where $\bm{x}_t$ is the current position and $a^\star_t$ is the randomly picked attenuation. We define as \textit{action} the next position to be taken when in state $s_t = \{\bm{x}_t,a^\star_t\}$ leading the drone to one of the positions in the set $\mathcal{X}_{a^\star_t}$.
Then, the policy $\pi(s_t)$ maps state $s_t$ to the next position, i.e.  
\begin{equation}
\pi(s_t) = \bm{x}_{t+1}.
\end{equation}
When passing from state $s_t$ to state $s_{t+1} = \{\bm{x}_{t+1},a^\star_{t+1}\}$, we define the {\em instantaneous reward} from \eqref{eq:energy} as
\begin{equation}
    R({s}_t,{s}_{t+1}) = -  \epsilon(\bm{x}_t,\bm{x}_{t+1}) \,.
\end{equation}
Note that the selection of a specific action at time $t$ is not enough to determine the state at $t+1$, since $a^\star_{t+1}$ is random and independent from $\bm{x}_{t+1}$.

Finally, we formally define the long-term optimization problem
\begin{equation}
  \{\pi^*(s)\} =   {\rm argmax}_{\{\pi(s)\}} E\left[\sum_{t=0}^{\infty}  R({s}_t,{s}_{t+1})\right]\,,
\end{equation}
where the expectation $E(\cdot)$ is taken over both the states and the actions.
%Indeed, a straightforward enumeration algorithm is unfeasible in practice its cost would grow exponentially with the number of possible actions to take and the number of forward steps to check. Thus, we consider three approaches to tackle this problem: the \ac{pg}, the \ac{bi}, and the STD-based heuristic solution.

\subsection{Bellman Iterative (BI) Solution}
%This however would require at least \footnote{i.e., assuming that only one value $\bm{x}^\star$ is associated with one $a^{\star}$} $\prod_{i = 1}^{|\mathcal{A}|} |\mathcal{X}_{a^\star_t}|$ operations.
To solve the \ac{mdp}, we resort to the Bellman equation and dynamic programming. In particular,  the {\em value update} function is %\hl{ho tolto $R_{\pi(s_t)}(s_t,s_{t+1})$ perche' non era definito e non serve}
%\begin{equation}\label{eq:valueUPD}
%    V(s_t) = \sum_{{s}_{t+1}} P_{\pi(s)}(s_t,s_{t+1})\left(R_{\pi(s)}(s_t,s_{t+1}) + \sum_{a \in \mathcal{A}}\gamma V(s_{t+1})\right)\,,
%\end{equation}
\begin{equation}\label{eq:valueUPD}
    V(s_t) = \frac{1}{|\mathcal{A}|}\sum_{\substack{{s}_{t+1}  = \\ \{\pi(s_t), a_{t+1}^*\}}} \underbrace{\left(R(s_t,s_{t+1}) + \sum_{a \in \mathcal{A}}\gamma V(s_{t+1})\right)}_{R'(s_t,s_{t+1})},
\end{equation}
and the corresponding {\em policy update} is 
%\begin{equation}\label{eq:policyUPD}
 %   \pi(s_t) = \argmax_v \sum_{{s}_{t+1}} P_{v}(s_t,s_{t+1})\left(R_{v}(s_t,s_{t+1}) + \sum_{a \in \mathcal{A}} \gamma V(s_{t+1})\right) ,
%\end{equation}
\begin{equation}\label{eq:policyUPD}
    \pi(s_t) = \frac{1}{|\mathcal{A}|}\argmax_{\bm{v}:a(\bm{v}) = a_t} \sum_{s_{t+1} = \{\bm{v}, a_{t+1}^*\}} R'(s_t,s_{t+1}) ,
\end{equation}
where $0<\gamma<1$ is the {\em discount factor}, chosen to balance greedy behaviors over later rewards.
To solve the problem we consider the Bellman iterative  (BI) solution \cite[Ch. 1]{Feinberg02Handbook}, where at each step $i$ we update the values as 
\begin{equation} \label{eq:rh_lh}
    V_{i+1}(s_t) = \frac{1}{|\mathcal{A}|} \max_{\bm{v}:a(\bm{v}) = a_t}   \sum_{  s_{t+1} = \{\pi(s_t), a_{t+1}^*\}}R'_i(s_t,s_{t+1}),
\end{equation}
where $R'_i(s_t,s_{t+1})$ is computed using \eqref{eq:valueUPD} with $V_i(s_{t+1})$. The procedure is repeated until either 1) a maximum predefined number of iterations is reached or 2) the right-hand side coincides with the left-hand side in \eqref{eq:rh_lh}.

In the worst-case scenario, the total computational cost is due to  the total number of states, and the number of possible actions. More in detail, we have $M=N |\mathcal{A}| $ states, and, at most, $ K \triangleq \max_{a^\star \in \mathcal{A}}|\mathcal{X}_{a^\star}|$ actions, leading to a total of  $|\mathcal{A}|  N^2   M  K $ computations in the worst case.

\subsection{Purely Greedy (PG) Solution}
The \ac{pg} aims at minimizing the long-term cost by maximizing the instantaneous reward, i.e., the one requiring less energy to be reached.
Given the position at time $t$, $\bm{x}_t$, the next position $\bm{x}_{t+1}$ is the closest in $\mathcal{X}_{a^\star_t}$. Formally, 
\begin{equation}\label{eq:greedy}
    \pi(s_t) = \argmin_{\bm{x}_{t+1}\in \mathcal{X}_{a^\star_t} } \epsilon(\bm{x}_t,\bm{x}_{t+1}) \,.
\end{equation}
While easy to implement, such a solution is in general not optimal as it is not always true that by maximizing the instantaneous reward we maximize also the long-term reward.

\subsection{Standard Deviation-based (STD-based) Solution}
We now propose another heuristic approach that estimates the \textit{strategic   value} $Y(\bm{x})$ of each position $\bm{x}\in \mathcal{X}$, such strategic value evaluates how convenient is for the drone to pick one position over the other, and replaces the $V(s)$ computed by the \ac{bi} solution. We thus have to design function $Y(\bm{x})$ and a proper policy.

%We want a position $\bm{x}$, to have a high value $Y(\bm{x})$, if it includes many different attenuations, and thus when in such region, the drone can reach the required attenuation with a low energy cost. 
Indeed, a position $\bm{x}$ should be associated with a high value $Y(\bm{x})$ if several different attenuations can be found in its neighborhood, therefore it is highly probable that when in $\bm{x}$, the drone will get the required attenuation with movements to nearby positions, thus saving energy.
To measure such diversity we consider the \ac{std} of attenuations in the neighborhood of each $\bm{x}$. In detail, let us introduce the set of $L\times L$ positions in $\mathcal X$ closest to (and including) $\bm{x}$ as $W_{\bm{x}}$. Next, we define the strategic value as  
\begin{equation}
    Y(\bm{x}) = \sqrt{\sum_{\bm{x}'\in W_{\bm{x}}}  \left( a(\bm{x}') - \mu_{\bm{x}} \right)^2 }\,, \quad \mu_{\bm{x}} = \frac{1}{L^2} \sum_{\bm{x}'\in W_{\bm{x}}}a(\bm{x}')\,.% \mathrm{STD}_{\bm{x}\in W(\bm{x})} \left( a(\bm{x})\right)\;.
\end{equation}
We remark that the computational cost to derive $Y(\bm{x})$ is $ N L^2  $, which is much lower than that of $V(s)$.

%where, with a slight abuse of notation, with $a\big( W(\bm{x})\big)$ we refer to (vector of) attenuations computed the window of positions in $W(\bm{x})$, centered in $\bm{x}$.  

Concerning the actual policy, while at the initial stage, getting the drone with the zone with high $Y(\bm{x})$ should have high priority, at later stages, we should already be in a zone with sufficiently high $Y(\bm{x})$, thus we could simply use the \ac{pg} strategy.
Thus, to balance such behaviors, at step $t$, we consider the following policy  
\begin{equation}\label{eq:heur}
    \pi(s_{t}) = \argmax_{\bm{x}_{t+1 } \in \mathcal{X}_{a^\star} }  \left\{ \delta e^{-t/\beta} Y(\bm{x}_{t+1})  -\epsilon(\bm{x}_t,\bm{x}_{t+1}) \right\} \:,
\end{equation}
where $\delta$ is the normalization parameter that allows to balance reward and \acp{std}, while $\beta$ is a decaying factor. Indeed, at later rounds, policies \eqref{eq:greedy} and \eqref{eq:heur} are equivalent.
The optimization in \eqref{eq:heur} can be done by enumeration over $\mathcal{X}_{a^*}$. % The decaying factor is chosen to make the heuristics more relevant for the first rounds than the latter, as for $n\gg \beta$, such strategy is equivalent to the purely greedy approach. 

\section{Numerical Results}\label{sec:results}
We report in this section the performance of the proposed protocol. 
We modeled a scenario where a drone moves around a plane area of size $\SI{50}{\meter} \times \SI{50}{\meter}$ at a height of $\SI{20}{\meter}$ from the center of the plane area where Alice is. The free space path loss is modeled via the Friis formula.
The area is sampled using a step size of $\SI{1}{\meter}$ along both directions, ending up with a total of $N = 2500$ sampled positions.
The considered carrier frequency is $f_0 = \SI{1.8}{\giga\hertz}$.
The shadowing coherence distance is $D_\mathrm{coh} = 10 \lambda$ with standard deviation $\sigma_{SH}=\SI{6}{\decibel}$, following the procedure reported in the Appendix. 
Fig.~\ref{fig:channel_sh_pl} shows a realization of the channel, which includes both free space path loss and shadowing.
To make the \ac{bi} solution computationally feasible, the quantizer ${\tt q}(\cdot)$ is uniform with 10 levels. 
As expected, the minimum attenuation is obtained approximately at the center, where the distance between Alice and Bob is at its minimum. On the other hand, the shadowing may cause both constructive and destructive interference:  for instance, we get a higher attenuation in the right than in the left corners of the map. 

\begin{figure}
    \centering
    \includegraphics[width=.95\columnwidth]{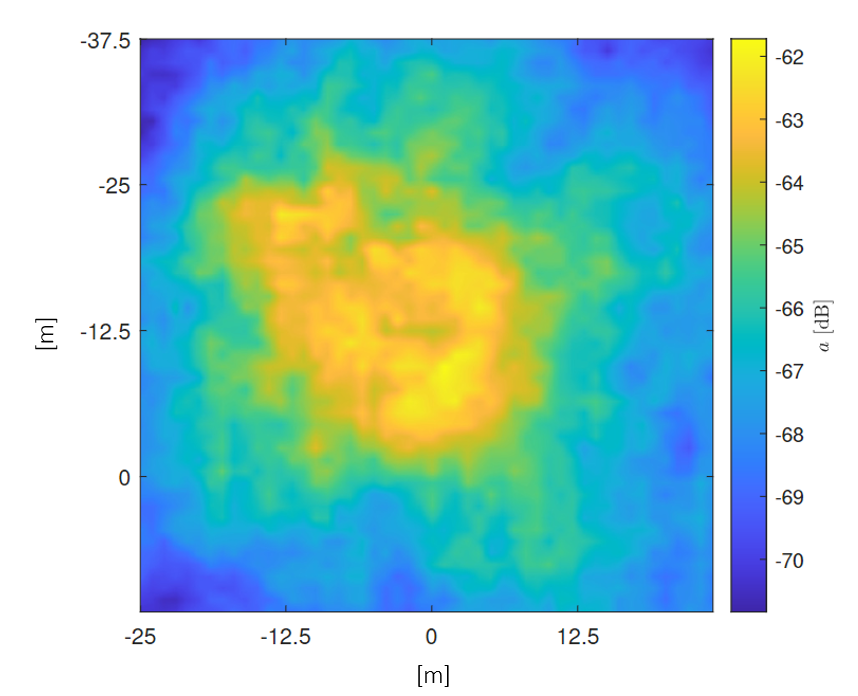}\caption{Example of channel realization, including both free space path-loss and shadowing over region $\mathcal{X}$.}\label{fig:channel_sh_pl}
    \label{fig:pl_shadowing}
    \vspace{-.4cm}
\end{figure}

\subsection{Security Performance}
\begin{figure}
    \centering
    \input{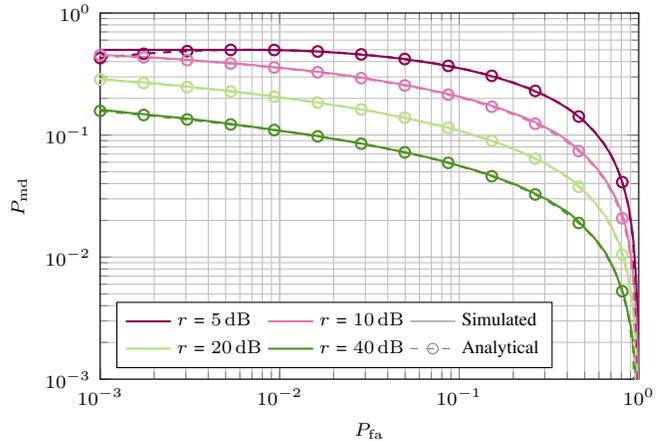}
    \caption{\ac{det} curves of the proposed authentication procedure for different $r$ values (in dB). Analytical (circles mark, dashed)  vs simulated (continuous).}
    \label{fig:detCurve}
    \vspace{-.4cm}
\end{figure}

First, we assess the security performance of the protocol.
Considering model \eqref{eq:hptest}, we set $r =5$, $10$, $20$, and $\SI{40}{\decibel}$.
Fig.~\ref{fig:detCurve} compares the \ac{det} curves obtained for different range values $r$ considering both simulation and the analytical derivation \eqref{eq:pmd}. We confirm the validity of our model, as the analytical model and simulation results almost perfectly match. As expected, when $r$ increases, it becomes harder for the attacker to guess the legitimiate $a^\star$, thus the $P_{\rm md}$ increases.

\subsection{Energy Optimization}

Next, we evaluate the results of the energy minimization policies described in Section~\ref{secopt}.
Concerning the parameters of such solutions, for \ac{bi} we fixed the discount factor $\gamma = 0.95$; for the \ac{std}-based solution, we consider instead windows $W(\bm{x})$ of size $L=5$, normalization factor $\delta = 100$, and decaying factor parameter $\beta= 20$.

\begin{figure}
    \centering
    \input{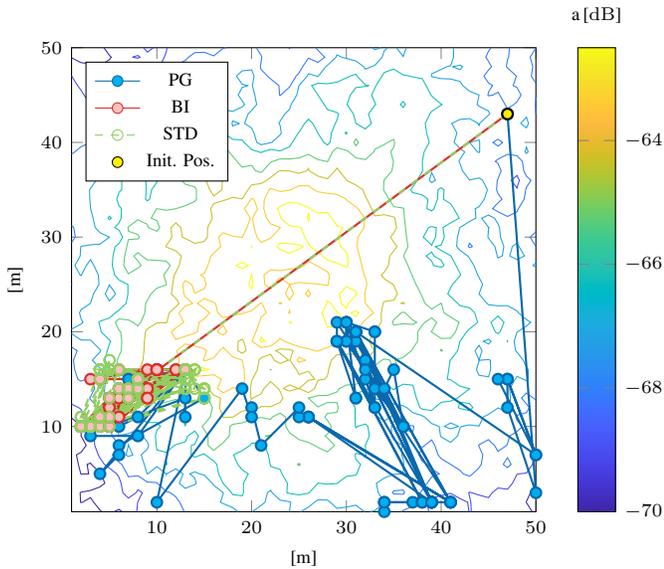} 
    \caption{Examples of trajectories obtained by using the \ac{pg} (blue), the \ac{bi} (red), and the STD-based approach (green). The contour plot shows $\hat{a}(\bar{\bm{x}})$ in dB.}
    \label{fig:trajExample}
    \vspace{-.4cm}
\end{figure}

Fig.~\ref{fig:trajExample} shows an example of 100 consecutive runs of the \ac{cr}-\ac{pla} protocol. All the energy minimization policies start from the same random position.
While all three strategies lead the drone to the same region, the \ac{pg} takes much longer to get the drone to that region, with it initially straying on the lower part of the map. Conversely, both \ac{bi} and \ac{std}-based solutions make the drones move a lot on the first step to get as soon as possible to the left corner. However, once there, the drone can stay locked in that advantageous region, with minimal movements. Thus, as designed, both strategies prioritize the long-term reward, at the expense of the instantaneous cost. The difference between \ac{bi} and the \ac{std}-based solutions is barely recognizable, with the latter often superimposing over the first solution. 
% \begin{figure}
%     \centering
%     \includegraphics[width=.95\columnwidth]{figures/path_loss.pdf}\caption{Results of the simulation for the attenuation due to path loss. Parameters from Tab.~\ref{tab:simParam}.}\label{fig:channel_pl}
%     \label{fig:pl_only}
% \end{figure}
%\begin{figure}
%    \centering
%    \subfloat[MDP-based (red) vs Purely Greedy (blue).\label{fig:tracj_new1}]{\input{figures/tracjetories_new1}}\\
%    \subfloat[STD-based (green) vs Purely Greedy (blue).\label{fig:tracj_new2}]{\input{figures/tracjetories_new2}} 
%    \caption{Examples of trajectories obtained by using the purely greedy, the MPD-based, and the STD-based approach. Parameters from Tab.~\ref{tab:simParam}.}
%    \label{fig:trajExample}
%\end{figure}

\begin{figure}
    \centering
    \input{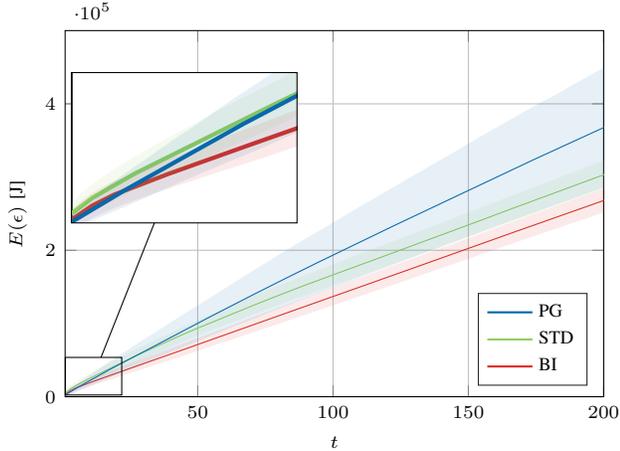}
    \caption{Average consumed energy $E[\epsilon(\bm{x}_t, \bm{x}_{t+1})]$ vs $t$, using the \ac{pg} (blue), the \ac{bi} (red), and the STD-based approach (green). Average (continuous line) plus/minus standard deviation (shaded area).}
    \label{fig:costRatio}
    \vspace{-.4cm}
\end{figure}

Fig.~\ref{fig:costRatio} shows the average consumed energy $E[\epsilon(\bm{x}_t, \bm{x}_{t+1})]$ as a function of $t$, after random initialization, obtained using the \ac{pg}, the \ac{bi}, and the STD-based approaches. As it can be seen from the magnification, on the first movements the greedy policy requires on average (slightly) less energy. However, after just a few steps, both \ac{bi}, and the STD-based solutions outperform the \ac{pg} solution, with a gap that increases over time.
As expected, the best performance is achieved by the \ac{bi}. Still, the gap between the optimal \ac{bi} and the \ac{std}-based approach is limited. Thus, in scenarios where the \ac{mdp}-based solution is not computationally feasible, it is reasonable to resort to the heuristic \ac{std} approach. 
%\begin{figure}
%    \centering
%    \input{figures/energy}
%    \caption{Average Energy Spent over time, using the greedy and the Bellman iterative approach.}
%    \label{fig:costRatio}
%\end{figure}

\section{Conclusions}\label{sec:conclusion}
We presented a novel \ac{cr}-\ac{pla} protocol for drone communication.
Using the previous knowledge of path loss, which includes modeling and sampling of free space path loss and shadowing, the drone manipulates the channel by moving on a specific position of the map. The verification procedure checks whether the measured path loss is consistent with the expected measurement. Additionally, the proposed protocol includes the long-term energy consumption in the design, while preserving the security of the system. Concerning the energy consumption minimization, three solutions have been considered: \ac{pg}, \ac{bi}, and \ac{std}-based. We compared the three solutions via numerical simulations, and a clear tradeoff between computational complexity and effectiveness in terms of energy consumed is observed.

\bibliography{biblio.bib}

\begin{thebibliography}{10}
\providecommand{\url}[1]{#1}
\csname url@samestyle\endcsname
\providecommand{\newblock}{\relax}
\providecommand{\bibinfo}[2]{#2}
\providecommand{\BIBentrySTDinterwordspacing}{\spaceskip=0pt\relax}
\providecommand{\BIBentryALTinterwordstretchfactor}{4}
\providecommand{\BIBentryALTinterwordspacing}{\spaceskip=\fontdimen2\font plus
\BIBentryALTinterwordstretchfactor\fontdimen3\font minus \fontdimen4\font\relax}
\providecommand{\BIBforeignlanguage}[2]{{%
\expandafter\ifx\csname l@#1\endcsname\relax
\typeout{** WARNING: IEEEtran.bst: No hyphenation pattern has been}%
\typeout{** loaded for the language `#1'. Using the pattern for}%
\typeout{** the default language instead.}%
\else
\language=\csname l@#1\endcsname
\fi
#2}}
\providecommand{\BIBdecl}{\relax}
\BIBdecl

\bibitem{Alam22}
M.~Alam, N.~Ahmed, R.~Matam, and F.~A. Barbhuiya, ``{IEEE} 802.11ah-enabled {I}nternet of drone architecture,'' \emph{IEEE Internet of Things Mag.}, vol.~5, no.~1, pp. 174--178, Mar. 2022.

\bibitem{Lin21}
X.~Lin, S.~Rommer, S.~Euler, E.~A. Yavuz, and R.~S. Karlsson, ``{5G} from space: An overview of {3GPP} non-terrestrial networks,'' \emph{IEEE Commun. Stand. Mag.}, vol.~5, no.~4, pp. 147--153, Oct. 2021.

\bibitem{ceccato21}
M.~Ceccato, F.~Formaggio, and S.~Tomasin, ``Spatial {GNSS} spoofing against drone swarms with multiple antennas and {W}iener filter,'' \emph{IEEE Trans. on Signal Proces.}, vol.~68, pp. 5782--5794, Oct. 2020.

\bibitem{michieletto22}
G.~Michieletto, F.~Formaggio, A.~Cenedese, and S.~Tomasin, ``Robust localization for secure navigation of {UAV} formations under {GNSS} spoofing attack,'' \emph{IEEE Trans. Autom. Sci. Eng}, pp. 1--14, Sept. 2022.

\bibitem{Adil23systematic}
M.~Adil, M.~A. Jan, Y.~Liu, H.~Abulkasim, A.~Farouk, and H.~Song, ``A systematic survey: {S}ecurity threats to {UAV}-aided {IoT} applications, taxonomy, current challenges and requirements with future research directions,'' \emph{IEEE Trans. Intell. Transp. Syst}, vol.~24, no.~2, pp. 1437--1455, Feb. 2023.

\bibitem{gupta2014cryptography}
P.~C. Gupta, \emph{Cryptography and network security}.\hskip 1em plus 0.5em minus 0.4em\relax PHI Learning, 2014.

\bibitem{CR-PLA:PCC}
S.~Tomasin, H.~Zhang, A.~Chorti, and H.~V. Poor, ``Challenge-response physical layer authentication over partially controllable channels,'' \emph{IEEE Commun. Mag.}, vol.~60, no.~12, pp. 138--144, Dec. 2022.

\bibitem{CR-PLA:DN}
F.~Mazzo, S.~Tomasin, H.~Zhang, A.~Chorti, and H.~V. Poor, ``Physical-layer challenge-response authentication for drone networks,'' in \emph{Proc. IEEE Global Communications Conf.}, 2023.

\bibitem{ACSA}
N.~Benvenuto, G.~Cherubini, and S.~Tomasin, \emph{Algorithms for Communications Systems and their Applications}, 2nd~ed.\hskip 1em plus 0.5em minus 0.4em\relax Wiley, 2021.

\bibitem{Abeywickrama18}
H.~V. Abeywickrama, B.~A. Jayawickrama, Y.~He, and E.~Dutkiewicz, ``Comprehensive energy consumption model for unmanned aerial vehicles, based on empirical studies of battery performance,'' \emph{IEEE Access}, vol.~6, pp. 58\,383--58\,394, Oct. 2018.

\bibitem{Feinberg02Handbook}
E.~Feinberg and A.~Shwartz, \emph{Handbook of Markov Decision Processes: Methods and Applications}.\hskip 1em plus 0.5em minus 0.4em\relax Springer, 2002.

\end{thebibliography}
\bibliographystyle{IEEEtran}

\balance

\appendix 

To model the shadowing effect we follow the procedure described in \cite[Sec. 4.1.10]{ACSA}. %Shadowing is a random variable that depends on both time and space. In this simplified model, a static channel with respect to time is considered, which means that shadowing across the space will be studied given a realization in time. The parameter that expresses the correlation between shadowing across space is the coherence distance, which is the basis for the model of shadowing spatial distribution in the following\cite{ACSA}.\newline
First, we consider a regular $N_1 \times N_2$ 2D grid with equal step size $A$, $\mathcal{X}$. Each point of the grid $(n_1,n_2)$ is mapped to a position $\bm{x}$, thus to and attenuation $\hat{a}(\bm{x}) = a(n_1,n_2)$. For simplicity, we consider $(0,0)$ to be the center of the grid. Then, the distance from the center of point $\bm{x}$ with coordinates $(n_1,n_2)$ is
\begin{equation}
    \Delta(\bm{x}) = A \sqrt{n_{1}^{2}+n_{2}^{2} } \,.
\end{equation}

%The complex Gaussian random variables will be the input of the filter that will relate points close to each other, with a correlation depending on the relative coherence distance, thus following the Gudmundson model of \eqref{eq:gudmundson}. Each sample in the grid is then related to others by the coherence distance, $D_{\rm coh}$.

Thus, considering the Gudmonson model for the autocorrelation function $r_{SH}(\Delta)$ (\ref{eq:gudmundson}), the associated power spectral density is
\begin{equation}
    \mathcal{P}(\bm{x}) = \mathrm{DFT}_2\left[r_{SH}\left(\Delta\right)\right] = \mathrm{DFT}_2\left[r_{SH}\left(A \sqrt{n_{1}^{2}+n_{2}^{2}}\right)\right] \,,
\end{equation} 
where $\mathrm{DFT}_2(\cdot)$ indicates the 2D discrete Fourier transform. 
Next, we generate a set of i.i.d. complex Gaussian samples, one for each point in the grid, $w\sim \mathcal{CN}(0,\,\sigma^{2}_{SH})$. %  is generated on the space $\mathcal{X}$ where sample is the noise sample associated with point $\bm{n}$.
Then, the shadowing component $a_{SH}(\bm{x})$ is obtained by the 2D convolution
\begin{equation}\label{eq:shadowing_convolved}
    a_{SH}(\bm{x}) = \sum_{\bm{x}'\in\mathcal{X}}{{h}_\mathrm{sh}(\bm{x}')\, w(\bm{x}-\bm{x}')}\;,
\end{equation}
where the 2D filter impulse response is obtained as 
\begin{equation}
    \mathcal{H}_{\rm sh}(\bm{x}) = \mathcal{K} \sqrt{\mathcal{P}(\bm{x})}\quad \mathrm{and} \quad h_{\rm sh}(\bm{x}) = \mathrm{DFT}_2^{-1}\left[\mathcal{H}_{\rm sh}(\bm{x})\right] \,,
\end{equation}
 where $\mathcal{K}$ is normalization coefficient chosen such that $h_\mathrm{sh}$ has unitary energy.
%\begin{equation}
%    \sum_{\bm{n}\in \mathcal{S}}{|h_\mathrm{sh}(\bm{n})|^2} = \frac{1}{N_{1}N_{2}}\sum_{\bm{k}\in \mathcal{S}}{|\mathcal{H}_\mathrm{sh}(\bm{k})|^2} = 1.
%\end{equation}

%This can be obtained by the normalization:
%\begin{equation}\label{eq:filter}
%    h_\mathrm{sh}(\bm{n}) = \frac{\hat{h}_\mathrm{sh}(\bm{n})}{\sqrt{\sum_{\bm{n}\in \mathcal{S}}{|\hat{h}_\mathrm{sh}(\bm{n})|^2}}}
%\end{equation}
%where
%\begin{equation}
%    \hat{h}_\mathrm{sh}(\bm{n}) = DFT_{2}^{-1}\left[\mathcal{\hat{H}}_\mathrm{sh}(\bm{k})\right] \quad \mathrm{and} \quad\mathcal{\hat{H}}_\mathrm{sh}(\bm{k})=\sqrt{\mathcal{P}(\bm{k})}
%\end{equation}

% Finally, the channel maps such as Fig.~\ref{fig:channel_sh_pl} are computed as in \eqref{eq:shadowed_pl}, neglecting the fading term and computing the free space path-loss using the  Friss formula.

\end{document}